\documentclass[apsrev,twocolumn,epsfig,tighten]{revtex4}
\usepackage[dvips]{graphicx}

\newcommand{\modulus}[1]{\vert {#1}\vert}
\newcommand{\ve}[1]{\mbox{\boldmath $#1$}}
\newcommand{\cprod}[2]{{#1}\!\times\!{#2}}

\begin{document}
\title{Dislocation-Mediated Melting in Superfluid Vortex Lattices}

\author{S. Andrew Gifford and Gordon Baym}
\affiliation{Department of Physics, University of Illinois at
Urbana-Champaign, 1110 West Green Street, Urbana, IL 61801}
\date{\today}

\begin{abstract}

    We describe thermal melting of the two-dimensional vortex lattice in a
rotating superfluid by generalizing the Halperin and Nelson theory of
dislocation-mediated melting. and derive a melting temperature proportional to
the renormalized shear modulus of the vortex lattice.  The rigid-body rotation
of the superfluid attenuates the effects of lattice compression on the energy
of dislocations and hence the melting temperature, while not affecting the
shearing.  Finally, we discuss dislocations and thermal melting in
inhomogeneous rapidly rotating Bose-Einstein condensates; we delineate a phase
diagram in the temperature -- rotation rate plane, and infer that the thermal
melting temperature should lie below the Bose-Einstein transition temperature.

\end{abstract}

\maketitle

\section{Introduction}
\label{sec:intro}

    Experiments on rapidly rotating Bose-Einstein condensates (BEC) in
harmonic magnetic traps have observed triangular lattices of vortices
\cite{abo,exp1,exp2,Dalibard}.  The lattice is expected to melt at
sufficiently rapid rotation rates, $\Omega$, close to the radial trap
frequency, $\omega_\perp$ \cite{Cooper, MacD, BaymT, GhoshBas}, at
sufficiently high energy temperature \cite{GiffBaym}.  In the rapidly rotating
regime, the condensate wave function is a linear superposition of single
particle states of the lowest Landau level (LLL) in the Coriolis force;
quantum fluctuations are predicted \cite{Cooper, MacD, BaymT, GhoshBas} to
melt the lattice when the number of vortices, $N_{\rm v}$, is so large that
the number of particles per vortex (or filling factor), $\nu$, falls to $\sim$
6-10.  While quantum fluctuations increase with decreasing $\nu$, thermal
fluctuations at finite temperature increase more rapidly.  The ratio of the
mean square vortex displacements due to thermal fluctuations to those due to
quantum fluctuations is $\sim 10^3\nu^{-{2/3}}(T/\Omega)\ln N_{\rm v}$
\cite{GiffBaym}.  Under typical experimental conditions, melting of the vortex
lattice induced by thermal fluctuations is more likely to be observed before
melting due to quantum fluctuations.

    In this paper, we generalize the Halperin and Nelson (HN) theory of
dislocation-mediated melting \cite{HalNel} to apply to the thermal melting of
vortex lattices.  The HN theory describes thermal melting of lattices as a
Kosterlitz-Thouless transition \cite{KosThou} due to unbinding of pairs of
elastic dislocations in the lattice.  The complication in generalizing the
theory is that the forces between vortices in neutral condensates are long
ranged, unlike the forces in a crystal lattice.  The HN theory has been
applied to vortex lattices in type-II superconductors \cite{dfish}, where the
forces between vortices are short ranged.  It has also been discussed in
superfluid helium films \cite{HubDon} under the very limiting assumption that
the lattice is incompressible.

    We consider a dilute gas of atoms trapped in an external potential and
rotating about the z-axis, taking a zero range s-wave interacation between
particles with coupling constant, $g=4\pi\hbar^2 a_{\rm s}/m$, where $a_{\rm
s}$ is the s-wave scattering length for two bosons; we work in the usual mean
field approximation for the interaction energy.  Furthermore, we describe the
elastic behavior of a vortex lattice in a rapidly rotating BEC using the
coarse-grained, long wavelength elastohydrodynamics description \cite{BC}.  In
lieu of treating the vortices as discrete entities, we express the deviation
of the vortices from their equilibrium positions (the triangular lattice
sites) as a continuous displacement field, $\ve{\epsilon}$.  Although the
condensates are three dimensional, the dominant motion is in the rotational
plane, to which we confine our attention.  The equations of motion couple the
local (coarse-grained) velocity field, $\ve{v}$, the local (coarse-grained)
particle density, and the vortex displacement field.  In
Refs.~\cite{BaymT,GiffBaym}, we used the elastohydrodynamic equations of
motion to derive the modes and various correlation functions of system.  Here,
we use the time-independent form of the same equations to understand the
elastic behavior and thermal melting of a superfluid vortex lattice.

    Dislocations are expected to form through nonlinear instabilities of
vortex lattice excitations, i.e., Tkachenko modes.  However, to determine the
melting, we need consider only the thermodynamics of dislocations.  In doing
so, we separate the elastic displacement field into two parts:  a fluctuating
part from the Tkachenko modes and a singular part from the dislocations.  In
the absence of nonlinearities these two terms are independent.

    Prior to deriving the thermal melting of a superfluid vortex lattice, we
first review the elastic equations for a superfluid vortex lattice in
Sec.~\ref{sec:equ}.  Then in Sec.~\ref{sec:dis}, we find the elastic
displacement field and energy of a single dislocation, and the interaction
energy of a dislocation pair.  In Sec.~\ref{sec:melt}, we generalize the HN
theory of two dimensional melting to a superfluid vortex lattice.  Finally, in
Sec.~\ref{sec:inhomo}, we consider melting in an inhomogeneous system.

\section{Elasticity}
\label{sec:equ}

    In this section we discuss the elastic properties of the vortex lattice,
working in the frame in corotating with the lattice.  Considering only motion
in the transverse plane, we write the energy as a function of the
two-dimensional density $n$, $\ve v$ and the vortex displacement field,
$\ve{\epsilon}$ as
\cite{elastic}
\begin{eqnarray}
\label{eden}
  E\{n,\ve{v},\ve{\epsilon}\} = && \int \left[\frac12mnv^2 +
  nV(\ve{r}) - \frac12mn\Omega^2r^2\right.\\ \nonumber
  &&\hspace{72pt}\left. + \frac12g_{\rm 2d}n^2 + \mathcal{E}_{\rm
el}\right]d^2r.
\end{eqnarray}
The first term in Eq.~(\ref{eden}) is the kinetic energy of the superfluid
surrounding the vortex lattice, in the second $V(\ve{r})$ is the trapping
potential, and third term is the centrifugal potential; the fourth term is the
interaction energy of the BEC, with $g_{\rm 2d}$ the usual two-dimensional
coupling constant ($\sim Zg$, where $Z$ is the axial thickness of the
condensate); the final term is the elastic energy density.  To second order in
$\epsilon$,
\begin{eqnarray}
\label{eel}
   \mathcal{E}_{\rm el} & =&
  2C_1({\ve\nabla}\cdot{\ve\epsilon})^2
  + C_2\left[\left(\frac{\partial
  \epsilon_x}{\partial x}-\frac{\partial\epsilon_y}{\partial y}\right)^2
  \right.\nonumber \\
   && \hspace{0.25 in}\left. + \left(\frac{\partial \epsilon_x}{\partial y}
   +\frac{\partial\epsilon_y}
    {\partial x}\right)^2\right] + \gamma n_{\rm v}
    \ve\epsilon \cdot\ve\nabla n.
\end{eqnarray}
The first two terms in Eq.~(\ref{eel}) are the compression and shear
energies, respectively, of the vortex lattice; $C_1$ and $C_2$ are the
(two-dimensional) density-dependent compressional and shear moduli
(essentially the usual three dimensional constants \cite{BC} integrated over
the thickness of the system in the axial direction) \cite{lame}.  In the
incompressible limit, e.g., in liquid helium, $C_2=-C_1=n\hbar\Omega/8$
\cite{tkachenko,BC}, while in the LLL limit ($g_{\rm 2D}n\ll \hbar\Omega $),
$C_2 \approx 0.119g_{\rm 2D}n^2$ \cite{MacD,sonin}, and $C_1$ is expected to
be small \cite{wbp2}.  The final term, in which $n^0_{\rm v} =
m\Omega/\pi\hbar$ is the equilibrium vortex density, is the coupling of the
vortex displacements to density inhomogeneities, with $\gamma$ an
$\Omega$-dependent constant \cite{elastic}.  In the incompressible limit,
$\gamma=(\pi\hbar^2/m)\ln(\ell/\xi_{\rm c})$, where $\ell = 1/\sqrt{\pi n_{\rm
v}}$ is the vortex (Wigner-Seitz) cell radius, $\ell^2 = \hbar/m\Omega$, and
$\xi_{\rm c}$ is the vortex core radius \cite{sheehy,elastic}; in the LLL
limit, $\gamma=\pi\hbar^2/m$ \cite{elastic}.

    We first calculate the response of the vortex lattice to a stationary
stress to linear order, assuming that the system is effectively two
dimensional with motions only in the plane perpendicular to the
rotational axis.  To linear order, thermal and quantum fluctuations
of the vortex positions do not influence the elastic response.  The
velocity field entering Eq.~(\ref{eden}) is governed, in a
stationary configuration, by the displacement field of the vortices.
Conservation of circulation implies that a shift of the positions of
the vortices from their equilibrium positions alters the local
velocity of the fluid according to \cite{BC}:
\begin{equation}
  \ve{v} +2\cprod{\ve{\Omega}}{\ve{\epsilon}} =
  \frac{\hbar}{m}\ve{\nabla}\Phi,
\label{veldisrel}
\end{equation}
where $\Phi$ is the non-singular superfluid phase.  The curl of this
equation, $(\cprod{\ve{\nabla}}{\ve{v}})_z =
-2\Omega\ve{\nabla}\cdot\ve{\epsilon}$, relates the transverse
component of the
velocity field to the longitudinal component of the displacement field.  The
steady state velocity field created by a distortion of the vortex lattice
obeys $\ve{\nabla}\cdot n\ve{v}=0$.  For a homogeneous system,
$\ve{\nabla}\cdot\ve{v}=0$, i.e., the longitudinal component of the velocity
vanishes.

    Variation of Eq.~(\ref{eden}) with respect the particle density and
displacement field gives two first order equations of motion,
\begin{equation}
\label{eom1}
  \left(\frac{\delta\mathcal{E}}{\delta n}\right)_{v,\epsilon}
  = -\gamma n_v^0 \ve\nabla\cdot\ve\epsilon  + g_{\rm 2D}n = \mu,
\end{equation}
where $\mu$ is the chemical potential; and
\begin{eqnarray}
\label{elforce}
  \left(\frac{\delta\mathcal{E}}{\delta\ve{\epsilon}}\right)_{n,v}
  &=& - 4\ve\nabla(C_1\ve\nabla\cdot\ve\epsilon) -
  2\ve{\nabla}\cdot(C_2\ve{\nabla})\ve\epsilon + \gamma n_v^0
\ve\nabla n
\nonumber \\
   &\equiv& \ve{\sigma} + \ve{\zeta}_n,
\end{eqnarray}
where $\ve{\sigma}$ is the elastic stress \cite{stressf}, and
$\ve{\zeta}_n = \gamma n_v^0 \ve\nabla n$ is the elastic stress due to density
inhomogeneity.  (Unlike in \cite{elastic} we separate out the $\zeta_n$ term
explicitly in defining $\sigma$.)  In general the elastic constants, dependent
on the density, appear within the derivatives.  We first consider a
homogeneous system, and return to the problem of an inhomogeneous system in
Sec.~\ref{sec:inhomo}.

    The fluid velocity induced by a displacement of the vortices results in a
Coriolis force which balances the elastic and external stresses, $\ve{\zeta}$
(including $\ve{\zeta}_n$):
\begin{equation}
\label{main1}
  2mn\cprod{\ve{\Omega}}{\ve{v}} = -\ve{\sigma} - \ve{\zeta}.
\end{equation}
The curl and divergence of Eq.\,(\ref{main1}) yield the transverse and
longitudinal elastic responses, $\epsilon_{\rm T}$ and $\epsilon_{\rm L}$,
independently.  Since $\ve{\nabla}\cdot n\ve{v}=0$,
\begin{equation}
\label{tterm}
  -\sigma_{\rm T} = 2C_2\nabla^2\epsilon_{\rm T} = \zeta_{\rm T},
\end{equation}
and
\begin{equation}
\label{lterm}
  -\sigma_{\rm L} - 2mn\Omega v_{\rm T} =
  [(4C_1+2C_2)\nabla^2 - 4\Omega^2nm]\epsilon_{\rm L} = \zeta_{\rm L}.
\end{equation}

    Fourier transforming Eqs.  (\ref{tterm}) and (\ref{lterm}), we find the
elastic Green's functions, $G_{ij}$ \cite{stable},
\begin{equation}
\label{greenll}
  G_{\rm LL}(\ve{k}) = \frac{\delta\epsilon_{\rm L}}{\delta\zeta_{\rm L}}
  = -\frac{1}{2(2C_1+C_2)k^2 + 4\Omega^2nm},
\end{equation}
\begin{equation}
\label{greentt}
  G_{\rm TT}(\ve{k}) = \frac{\delta\epsilon_{\rm T}}{\delta\zeta_{\rm T}}
  = -\frac{1}{2C_2 k^2},
\end{equation}
and
\begin{equation}
\label{greenlt}
  G_{\rm LT}(\ve{k}) = G_{\rm TL}(\ve{k}) = 0;
\end{equation}
transverse stresses give rise only to transverse displacements, and
longitudinal stresses only to longitudinal displacements.  In terms of the
elastic Green's functions, the dynamic equations for displacement field are
\cite{BaymT},
\begin{eqnarray}
 i\omega\epsilon_{\rm T} + \frac{1}{2\Omega mn}G_{\rm LL}^{-1}(\ve{k})
 \epsilon_{\rm L} &=& \frac{\zeta_{\rm L}}{2\Omega mn}, \nonumber \\
   i\omega\epsilon_{\rm L} +
   \left(\frac{2\Omega \omega^2}{\omega^2 - s^2 k^2}
   -\frac{1}{2\Omega mn}  G_{\rm TT}^{-1}(\ve{k}) \right)\epsilon_{\rm T}
    &=& -\frac{\zeta_{\rm T}}{2\Omega mn}.
   \nonumber\\
  \label{gfe}
\end{eqnarray}

    Unlike in conventional elastic systems, the longitudinal elastic Green's
function contains an additional length scale, an elastic ``penetration depth":
\begin{equation}
   \delta^{2} = \frac{2C_1+C_2}{2\Omega^2mn}.
\end{equation}
Note that $C_2>0$ while $C_1$ is generally negative \cite{wbp}, and thus
$\delta^2$ can be positive or negative.  In the incompressible regime
\cite{delta}, $\delta^2 = -\ell^2/16$, while in the LLL regime, $\delta^2
\approx 0.03\ell^4/\xi^2 = 0.06(g_{\rm 2D}n/\hbar \Omega)\ell^2$, where $\xi^2
= \hbar^2/2g_{\rm 2D}nm$ is the healing length in the condensate.  In the LLL
regime, $\hbar\Omega \gg g_{\rm 2D}n$, and hence in both regimes
$\modulus{\delta}\ll \ell$.  Thus within the regime of validity of the
coarse-grained continuum approximation, $G_{\rm LL}(\ve{k}) \simeq
-1/4\Omega^2nm$, and $G_{\rm LL}(\ve{r},\ve{r}') \approx
(-1/4\Omega^2nm)\delta(\ve{r}-\ve{r}')$.

    The elastic displacement due to the (longitudinal) driving force
$\ve{\zeta}_n = \gamma n_v^0 \ve\nabla n$ is thus
\begin{eqnarray}
  \ve \epsilon(\ve{r})
  &=& \int G_{LL}(\ve{r},\ve{r}')\zeta_n(\ve{r}')d^2r'
  \nonumber\\
   &\approx&  -\frac{\gamma n_v^0}{4\Omega^2m}\ve\nabla \ln n(\ve{r}),
\label{rad-dis}
\end{eqnarray}
consistent with the results of Refs.~\cite{sheehy,elastic,wbp}.
Corrections to this result using the elastic Green's functions for an
inhomogeneous system are of order $1/N_{\rm v}$.

\section{Dislocations}
\label{sec:dis}

    To describe dislocation-mediated melting of a superfluid vortex lattice,
we first investigate the structure of elastic dislocations, which are
topological defects in the vortex lattice.  Such dislocations obey the
condition,
\begin{equation}
\label{burgerscond}
\oint d\ve{\epsilon} = \ve{b},
\end{equation}
for any closed path around the dislocation, where $\ve{b}$ is the Burger's
vector \cite{Landau, Nab}.  The minimum length of a Burger's vector is
the lattice spacing.

    We first solve Eqs.~(\ref{veldisrel}) and (\ref{main1}) for an elastic
dislocation in a superfluid vortex lattice, leaving mathematical details for
the Appendix.  For $\ve{b}=b\ve{\hat{x}}$ and $\delta=0$, the displacement
field is,
\begin{eqnarray}
\label{exbxd0}
\nonumber
\epsilon_x &=&
\frac{b}{2\pi}\left[\tan^{-1}\left(\frac{y}{x}\right) - \frac{xy}{r^2} +
\frac{2C_2}{\Omega^2mn}\frac{xy}{(x^2+y^2)^2}\right]
\\
\nonumber
\label{eybxd0}
\epsilon_y &=& \frac{b}{2\pi}\left[\frac12\ln\left(x^2+y^2\right) +
\frac{x^2}{r^2} + \frac{C_2}{\Omega^2mn}\frac{y^2-x^2}{(x^2+y^2)^2}\right].
\nonumber
\\
\end{eqnarray}
This displacement field is transverse, with vanishing divergence away from
the origin, as expected since the longitudinal Green's function vanishes away
from the dislocation.

     In addition we find that the energy of a single dislocation is
\begin{widetext}
\begin{equation}
\label{sdenergy}
   E_{\rm dis} = \frac{C_2b^2}{\pi}\left[\ln\left(\frac{R}{a_{\rm c}}\right) +
   \frac{4C_2\delta^2}{2C_1+C_2}\left[\frac{1}{a_{\rm
   c}^2}\left(1-\frac{a_{\rm
   c}}{2\delta}K_1\left(\frac{a_{\rm c}}{\delta}\right)\right) -
  \frac{1}{R^2}\left(1-\frac{R}{2\delta}K_1
  \left(\frac{R}{\delta}\right)\right)\right]\right],
\end{equation}
\end{widetext}
where $K_n$ are modified Bessel functions of the second kind, and
$a_{\rm c}$ is the radius of the dislocation core; $a_{\rm c}>a$, the lattice
spacing.  We include the dislocation core energy in the chemical potential of
the system.  In the limit $R\to\infty$, the logarithmic term in
Eq.~(\ref{sdenergy}) is the dominant term.

    To second order in the displacement field the interactions between
dislocations are pairwise.  The energy of two separated dislocations at
positions $\ve{r}_{\alpha}$ and $\ve{r}_{\alpha'}$, with Burger's vectors
$\ve{b}_{\alpha},\ve{b}_{\alpha'}$, is, as derived in the Appendix,
\begin{widetext}
\begin{eqnarray}
\label{interdisE}
   E_{\alpha,\alpha'} &=& \frac{C_2}{\pi}
  \Bigg[\ve{b}_{\alpha}
 \cdot\ve{b}_{\alpha'}\ln\left(\frac{\modulus{\ve{r}_{\alpha}
 -\ve{r}_{\alpha'}}}{a_{\rm c}}\right) -
 \frac{\ve{b}_{\alpha}\cdot(\ve{r}_{\alpha}-\ve{r}_{\alpha'})
 \,\ve{b}_{\alpha'}\cdot(\ve{r}_{\alpha}-\ve{r}_{\alpha'})}
 {\modulus{\ve{r}_{\alpha}-\ve{r}_{\alpha'}}^2}
 \\
 && \hspace{18pt} +\varepsilon_{ij}\varepsilon_{kl}b_{\alpha,i}b_{\alpha',k}
 \frac{4C_2\delta^2}{(2C_1+C_2)}\frac{\partial^2} {\partial
 x_{\alpha',j}\partial x_{\alpha',l}}
 \left(\ln\modulus{\ve{r}_{\alpha}-\ve{r}_{\alpha'}}  +
 K_0\left(\frac{\modulus{\ve{r}_{\alpha}-\ve{r}_{\alpha'}}}
 {\delta}\right)\right)\Bigg],
 \nonumber
\end{eqnarray}
\end{widetext}
where $\varepsilon_{ij}$ is the two dimensional anti-symmetric tensor, The
first term in Eq.\,(\ref{interdisE}) grows with increasing separation of the
dislocation pairs and is therefore long ranged; the final term decreases with
the separation at a rate depending on $\delta$.  In the limit
$\delta\to\infty$, where one recovers classical elasticity, the final term
extend to long distances; in this limit, Eq.\,(\ref{interdisE}) reduces to the
dislocation interaction energy of Ref.~\cite{HalNel}.

    Two dislocations with opposite Burger's vectors are attracted with an
interaction which grows logarithmically.  Dislocations thus bind in pairs at
low temperature.  Dislocation-mediated melting is driven, as we discuss in the
following section, by unbinding of such pairs.

    The core radius, $a_{\rm c}$, the length scale at which the dislocation
can no longer be described by a {\it continuous} displacement field, as well
as the dislocation core energy, can only be computed from the local
microscopic condensate wave function.  In the incompressible limit, the energy
of the vortex lattice depends entirely on the kinetic energy of the
surrounding superfluid, and thus the core energy is linearly proportional to
the density.  Similarly, in the rotating frame in the LLL limit, only the
interaction energy changes with vortex position, and thus the core energy is
proportional to the square of the density.  We approximate the dislocation
core energy by,
\begin{equation}
  E_{\rm core} \approx a_1n + a_2n^2,
\end{equation}
where $a_1$ and $a_2$ are positive functions of the rotation rate, with
the approximate dependence, $a_1\propto \hbar\Omega a_{\rm c}^2$ and $a_2
\propto ga_{\rm c}^2$.  Since $E_{\rm c}$ depends on the position of the
vortices within the dislocation core, where the continuum limit is not valid,
exact values of $a_1$ and $a_2$ must be calculated in terms of the local
condensate wave function.

\section{Dislocation-Mediated Melting}
\label{sec:melt}

    The dislocation-mediated melting of the vortex lattice is a
Kosterlitz-Thouless transition \cite{KosThou} arising from unbinding of pairs
of elastic dislocations.  Thermal fluctuations cause dislocations to enter
large systems in pairs with opposite Burger's vectors.  Below the melting
temperature, pairs are bound; pairs of Burger's vectors of equal magnitude but
opposite direction produce no long range elastic deformation, and are thus
most stable than isolated dislocations.  If two opposite dislocations overlap,
they mutually annihilate.

    Above the melting temperature, dislocations are unbound, and to a first
approximation can be treated independently.  In the thermodynamic limit
($R\to\infty$), the energy (\ref{sdenergy}) is minimized for a dislocation
with the minimum magnitude of the Burger's vector, the lattice spacing, $a$,
and thus $E_{\rm dis} \simeq (C_2a^2/\pi)\ln(R/a_{\rm c})$ plus a small
constant.  [For a triangular lattice, $a^2=(2/\sqrt{3})/n_{\rm v}$.] This
logarithmic form allows us to apply the simple physical argument of Kosterlitz
and Thouless to determine when dislocations are thermodynamically stable in
the vortex lattice.  Since there are $\sim(R/a_{\rm c})^2$ possible locations
for a single dislocation in the vortex lattice, the entropy of a single
dislocation is $\sim 2\ln(R/a_{\rm c})$.  Hence, the Helmholtz free energy per
dislocation is
\begin{equation}
\label{disthermo}
  F \approx \left(\frac{C_2a^2}{\pi} - 2T\right)\ln\left(\frac{R}{a_{\rm
   c}}\right).
\end{equation}
Thus for $T > C_2a^2/2\pi \equiv T_{\rm m}^0$, dislocations (of size
$b=a$) are thermodynamically stable.  Isolated dislocations remain in the
system only above this temperature.

    In the absence of interactions between dislocation pairs, $T_{\rm m}^0$
would be the melting temperature.  As the temperature increases to $T_{\rm
m}^0$ the mean separation between two dislocations within a pair diverges; in
general the mean separation of a pair ${\alpha,\alpha'}$ is \cite{KosThou},
\begin{equation}
\label{rmsdis}
    \langle (\ve{r}_{\alpha} -\ve{r}_{\alpha'})^2 \rangle \propto \int
  e^{-\beta E_{\alpha,\alpha'}(\ve{r})}r^2 d^2r.
\end{equation}
For $T < T_{\rm m}^0$ the mean separation is finite, while for $T>T_{\rm
m}^0$, the logarithmic term in Eq.~(\ref{interdisE}) leads to a divergent
separation length within a dislocation pair, indicating that the pair
has become unbound.

    To obtain a more accurate description of melting requires considering
interactions of dislocation pairs, which can be taken into account by the
renormalization-by-decimation technique in Refs.~\cite{Kost, young}.  This
technique requires $\delta/R$ to be small.  We do not give the details here,
but note that the net effect is to renormalize the shear modulus $C_2$ to a
temperature dependent function, $C_{2,{\rm R}}(T)$.  The shear modulus has an
intrinsic dependence on the temperature, since it is proportional to the
superfluid density.  The melting temperature is then
\cite{deltadep}
\begin{equation}
\label{Tmeltd0}
  T_{\rm m} = \frac{a^2}{2\pi}C_{2,{\rm R}}(T_{\rm m}).
\end{equation}
Previous estimates for $C_{2,{\rm R}}$ for a lattice of particles
interacting logarithmically (e.g., vortices) have shown $C_{2,{\rm R}} \approx
AC_{2}$ where $A\lesssim 1$ \cite{dfish}.

    The melting temperature depends only on the shear modulus.  It is
instructive to compare the origin of this dependence in a system of vortices
with the corresponding result in a traditional elastic system (e.g., an
isotropic crystal) \cite{HalNel}.  The first two terms of
Eq.~(\ref{interdisE}) have the same form for the interaction energy of
dislocations in an elastic system.  The third term of Eq.~(\ref{interdisE}),
containing an additional length scale, $\delta$, is particular to the vortex
system; this term is a short range interaction, $\sim\delta^2/r^2$.  In
smectic liquid crystals near the nematic-to-smectic-A transition, the
interaction energy of two dislocations also contains an additional length
scale, a penetration depth; when this length is small enough to screen the
dislocations in the liquid crystal beyond their core, the dislocation pairs
unbind at $T=0$ \cite{Lub}.  In such a liquid crystal, \emph{all} the energy
channels (splay, twist, and bend) are screened by the compression or
stretching of the smectic layers.  In the superfluid vortex lattice, however,
only compressional effects are screened (by the rigid body motion) while the
shearing is unaffected.  In the vortex system the unbinding of the
dislocations is promoted by shearing alone.

    For temperatures approaching $T_{\rm m}$ from below, the displacement
correlation function is independent of the system parameters \cite{BaymT}:
\begin{equation}
\lim_{T\to T_{\rm m}^-} \frac{\left<\modulus{\ve{\epsilon\,}
(\ve{r}\,)-\ve{\epsilon\,}(\ve{r}\,')}^{2}\right>}{\ell^2}
\approx\frac{1}{2\sqrt{3}\pi}
\ln\left(\frac{\modulus{\ve{r}-\ve{r}'}}{\ell}\right).
\end{equation}
In the solid phase, the vortex lattice exhibits algebraic long-range
order.  However, above $T_{\rm m}$, the displacement correlations function
decay roughly exponentially with distance, as one sees from renormalization
group calculations \cite{HalNel}.

    Extended two dimensional Bose systems, e.g., helium films, or atoms in
optical lattices \cite{Dalibard2,cornell2,markus}, undergo a (conventional)
Kosterlitz-Thouless transition \cite{Bere,KosThou,NelKost} to a superfluid
state at $T_{\rm KT} = \pi\hbar^2n_{\rm s}/2m$, where $mn_s \equiv \rho_s$ is
the two-dimensional superfluid mass density.  When rotating at low
temperature, such a system should contain a vortex lattice, which melts via a
dislocation mediated Kosterlitz-Thouless transition at a temperature $T_{\rm
m} = a^2C_{2,{\rm R}}/2\pi$.  Using the incompressible shear modulus in
$T_{\rm m}$ one finds $T_{\rm m}/T_{\rm KT} = (1/4\pi\sqrt3)n_{\rm s}(T_{\rm
m})/n_{\rm s}(T_{\rm KT})$ \cite{HubDon}, where $n_{\rm s}(T_{\rm m})/n_{\rm
s}(T_{\rm KT}) \approx 1 + b(1-T_{\rm m}/T_{\rm KT})^{1/2}$; $b$ in $^4$He
films is measured to be $\simeq 5.5$ \cite{BisRep}.  Thus $T_{\rm m}/T_{\rm
KT} \approx 0.26$; renormalization of the shear modulus should reduce this
result at most by a factor of $2$ \cite{dfish}.

    Cozzini et al.~find from numerical computation of the $T=0$ shear modulus
vs.  $\hbar\Omega/g_{\rm 2D}n$ -- from the incompressible regime
($\hbar\Omega/g_{\rm 2D}n<<1$) to the LLL regime ($\hbar\Omega/g_{\rm
2D}n_s>>1$) -- that with the exception of a small increase just above the
incompressible regime, $C_2\lesssim n\hbar\Omega/8$ \cite{Cozzini}.  This
result suggests that $T_{\rm m}/T_{\rm KT}$ decreases with increasing
$\hbar\Omega/g_{\rm 2D}n_s$.  Preliminary calculations for the parameters of
the $^{87}$Rb experiment of Ref.~\cite{Dalibard2D} give $b\simeq 4.5$
\cite{Markus2,Markus3}, indicating $T_{\rm m}/T_{\rm KT} \lesssim 0.23$ prior
to renormalization effects.

\section{Melting in an Inhomogeneous System}
\label{sec:inhomo}

    We expect thermal melting in a rapidly rotating BEC in a finite
inhomogeneous geometry also to be driven by formation of dislocations arising
from thermal fluctuations.  Dislocations have in fact been observed in the
lattices in these systems; see Fig.~5a of Ref~\cite{abo}, However, describing
the structure and unbinding of dislocation pairs is much more difficult than
in a homogeneous system.  On the one hand, the elastic constants, $C_1$ and
$C_2$, depend on the local density, and dislocation energies depend on
position, while the elastic theory contains second order terms explicitly
dependent on $\ve{\epsilon}$ in addition to the usual terms dependent on
derivatives of $\ve{\epsilon}$ \cite{wbp2}.  Furthermore, in a finite system
dislocations can be created not only in pairs in the bulk of the system, but
an isolated dislocation can also be formed at the edge, which separates from
its image dislocation and moves into the bulk.  Such additional features must
be taken into account numerically to determine the details of melting, a
problem we leave for the future \cite{extdis}.

    Here we simply determine an upper bound on the melting temperature.  We
assume that the maximum density, $n_0$, of the BEC occurs at the center of the
trap, where a dislocation has maximum energy.  Furthermore, this maximum
energy is less than the energy of a dislocation in a condensate of the same
size but with uniform density, $n_0$.  As a dislocation approaches the edge of
the trap, its energy goes to zero.  Thus the dislocation energy in an
inhomogeneous system is bounded above by
\begin{equation}
    E_{\rm dis}(\ve{r}) \le \max\{E_{\rm dis}\} < (C_{2}(0)a^2/\pi)\ln(R_{\rm
    TF}/a_{\rm c}),
\end{equation}
where $C_{2}(0)$ is the (unrenormalized) shear modulus at the center of
the trap, and $R_{\rm TF}$ is the Thomas-Fermi radius of the rotating cloud.
Using an upper bound to the dislocation energy should yield an upper bound on
the melting temperature, and thus Eq.~(\ref{disthermo}) indicates that the
temperature at which dislocations are thermodynamically stable is bounded
above by $T^0_{\rm m} = C_{2}(0)a^2/2\pi$.  In fact, dislocations should
appear near the edge of the BEC for temperatures below $T^0_{\rm m}$ and creep
towards the center as the $T$ approaches $T^0_{\rm m}$, while for $T \gtrsim
T^0_{\rm m}$, we expect dislocations to occur throughout the system.

    Melting generally occurs below the Bose-Einstein condensation temperature,
$T_{\rm c} \approx 0.94\hbar(N\omega_z(\omega_{\perp}^2-\Omega^2))^{1/3}$,
where $\omega_z$ and $\omega_\perp$ are the azimuthal and radial angular
frequencies of the trap respectively.  We first estimate $T^0_{\rm m}/T_{\rm
c}$ in the limit of a nearly incompressible fluid ($\hbar\Omega \ll gn$) in
trap geometry, and then consider the LLL limit.  In the former regime,
$T^0_{\rm m} = C_{2}(0)a^2/2\pi \lesssim \hbar \Omega n_s(0) a^2/16\pi =
\rho_s(0)\hbar^2/8\sqrt3 m^2$, where $\rho_s(0)$ the (two dimensional)
superfluid mass density at the center of the trap.  The shear modulus depends
on the superfluid mass density a function of temperature, which for a weakly
interacting Bose gas is approximately the product of the mass, the mean
thickness, and the three dimensional condensate density, $n_{\rm TF}(r=0,T)$.
At the center of the trap the condensate density, coarse-grained over the
vortices, is approximately homogeneous and does not feel the effect of the
condensate edge; at $T=0$,
\begin{equation}
\label{TF_den}
  n_{\rm TF}(r=0,T=0) =
 \frac{\hbar\omega_\perp}{2gb_{\Delta}}
 \left[\frac{15Nb_{\Delta}a_s}{d_{\perp}}
 \frac{\omega_z}{\omega_\perp}\left(1-\frac{\Omega^2}{\omega_\perp^2}
 \right)\right]^\frac25.
\end{equation}
In writing Eq.~(\ref{TF_den}), we approximate the number of particles in
the condensate by the total particle number $N$;
$d_{\perp}=\hbar/\sqrt{m\omega_{\perp}}$ is the radial trap length, and
$b_{\Delta}\approx 1.1596$ is the Abrikosov parameter \cite{fischer}.  At
finite temperature, $n_{\rm TF}(r=0,T) = n_{\rm TF}(r=0,0)(1-(T/T_{\rm
c})^{3/2})$.  The thickness, $Z$, of the condensate at the center of the trap
is approximately $d_z (2gb_{\Delta}n_{\rm TF}(0)/\hbar\omega_z)^{1/2}$, where
$d_z = \hbar/\sqrt{m\omega_z}$ is the azimuthal trap length.

    Thus we find $T_{\rm m}/T_{\rm c}$ as the solution of
\begin{widetext}
\begin{equation}
\label{inhomo-melt-cond}
  \frac{T_{\rm m}}{T_{\rm c}}
  \lesssim 0.0049 N^{4/15}\left(\frac{d_z}{b_{\Delta}a_s}\right)^{2/5}
  \left(\frac{d_z}{d_{\perp}}\right)^{16/15}
  \left(1-\frac{\Omega^2}{\omega_\perp^2}\right)^{4/15}
  \left(1-\left(\frac{T_{\rm m}}{T_{\rm c}}\right)^{\frac32}\right).
\end{equation}
\end{widetext}
For the experimental parameters of Refs.~\cite{exp1,exp2} in $^{87}$Rb
($a_s = 4.8\pm0.05$~nm, $d_{\perp}=3.75$~$\mu$m, $d_z=4.69$~$\mu$m) with
$10^4$ particles and $\Omega=0.9\omega_{\perp}$, we have $T_{\rm m}/T_{\rm
c}\lesssim 0.49$.  For $N=10^6$ this ratio increases to $T_{\rm m}/T_{\rm
c}\lesssim 0.78$.  If $T_{\rm m}/T_{\rm c}$ is only slightly less than unity,
then the thermal melting of the vortex lattice will be hard to observe
experimentally.  Higher rotation rates help to lower the melting temperature,
but in most experiments, the number particles in the system decreases too,
making thermal melting harder to resolve.  Note that
Eq.~(\ref{inhomo-melt-cond}) predicts that $T_{\rm m}/T_{\rm c}$ decreases
with increasing scattering length, allowing for the possibility of inducing
thermal melting via a Feshbach resonance.

    In the LLL limit, the density at the center of the trap is \cite{baym2},
\begin{equation}
n_{\rm LLL}(r=0,T=0)
= \frac{1}{2\pi^{5/4}}\left(\frac{N}{b_{\Delta}d_{\perp}^4d_za_s}
\left(1-\frac{\Omega^2}{\omega_{\perp}^2}\right)\right)^{\frac12}.
\label{LLLden}
\end{equation}

    Along the axis, the condensate particles are in the lowest harmonic
oscillator state, so the thickness of the condensate at the center of the trap
is approximately $d_z$.  Also at the center of the trap, the condensate is in
a harmonic oscillator state along the axis and approximately homogeneous in
the plane of rotation, so the density as a function of temperature is
proportional to $1-(T/T_{\rm c})^2$, which we take as the temperature
dependence of $C_2$ in the LLL regime.  The melting temperature is given by
the solution of
\begin{widetext}
\begin{equation}
  \frac{T_{\rm m}}{T_{\rm c}} \approx
  0.0036\left(\frac{Nd_z}{d_{\perp}}\right)^{2/3}
  \frac{\omega_{\perp}}{\Omega}
  \left(1-\frac{\Omega^2}{\omega_{\perp}^2}\right)^{2/3}
  \left(1-\left(\frac{T_{\rm m}}{T_{\rm c}}\right)^{2}\right)^2.
\label{TmLLL}
\end{equation}
\end{widetext}

\begin{figure}[htbp]
\begin{center}
{\includegraphics[width=3.0in]{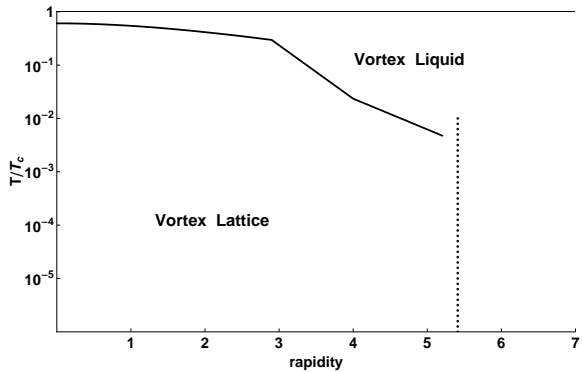}}
\caption{Sketch of the phase diagram of rapidly rotating BEC.  The
temperature is in units of the condensation temperature, $T_{\rm c}$; the
``rotational rapidity" is defined as $y= \tanh^{-1}(\Omega/\omega_{\perp})$
\cite{baym2}. To the right of the dotted line, the system at lower
temperatures enters a strongly correlated regime.
}
\label{fig:rrphase}
\end{center}
\end{figure}

    Figure~\ref{fig:rrphase} shows schematically the phase diagram of a
rapidly rotating BEC.  The y-axis is the temperature divided by the
condensation temperature, and the x-axis is the ``rotational rapidity," $y =
\tanh^{-1}(\Omega/\omega_{\perp})$ \cite{baym2}.  In this figure we use the
parameters of Refs.~\cite{exp1,exp2} as above.  The curve separating the
vortex lattice and vortex liquid phases is given by
Eq.~(\ref{inhomo-melt-cond}) at low rotation, (\ref{TmLLL}) at high
rotation rates, and an interpolation between the two limits; the two kinks in the curve are artifacts of using these two limits with the interpolation.  
We assume that the melting leads directly to a vortex liquid,
neglecting a possible hexatic phase intermediate between the vortex lattice
and liquid which exhibits no long-range order of the vortex positions but
maintains algebraic long-range order of the triangular orientation of the
lattice \cite{HalNel}.  Computing the phase boundary between such a hexatic
phase and a vortex liquid requires microscopic details of the dislocations
(e.g., core radius and energy) beyond the scope of this paper.

    The vortex lattice is predicted to melt via quantum fluctuations into a
new strongly correlated regime.  Details of this regime are discussed in
Refs.~\cite{Cooper,MacD,BaymLT}.  With the density Eq.~(\ref{LLLden}) and
$\nu\approx n/n_{\rm v}$, we estimate the phase boundary for the strongly
correlated system to be at
\begin{equation}
 \frac{\Omega_{\rm c}}{\omega_{\perp}} \approx
 \left[1+ \left(\frac{4\sqrt{\pi}b_{\Delta}a_{\rm s}}{Nd_z}\right)\nu_{\rm
  c}^2\right]^{-\frac12}.
\end{equation}
In Fig.~\ref{fig:rrphase}, where we take a critical filling fraction,
$\nu_{\rm c}\sim 10$ \cite{MacD,BaymT}, the phase boundary is indicated by the
vertical dotted line, at $y\simeq5.4$ (corresponding to $\Omega/\omega_{\perp}
\approx 0.99996$), The phase diagram remains to be studied in detail, e.g,
near the region where the vortex liquid phase boundary intercepts the strongly
correlated regime, and the behavior at higher temperatures.

    At low rotation rates ($\Omega/\omega_{\perp}\simeq 0.9$), the melting
temperature may be too close to the condensation temperature to discern the
melting of the vortex lattice experimentally.  Higher rotation rates provide a
better opportunity to observe melting since $T_{\rm m}$ decreases with
$\Omega/\omega_{\perp}$.  However, at sufficiently high rotation rates (e.g.,
at $y\sim 5$ in Fig.~\ref{fig:rrphase}), it is not clear how to experimentally
distinguish thermal melting from quantum melting, since we do not have an
adequate understanding of the melted states at finite temperature.

\section{Conclusion}

    The elasticity of a superfluid vortex lattice is determined by the energy
of the vortex cells and the global rotational fluid flow; the kinetic energy
of the ''rigid-body" rotational velocity stabilizes the system against a
negative compression modulus, $C_1$.  Because of the rotational motion, the
effect of elastic compression is significantly weakened, and shearing is the
dominant degree of freedom for the vortex lattice.  Correspondingly, we find
that the melting temperature of the vortex lattice depends only on the
(renormalized) shear modulus.

    Among the unanswered questions for future examination are, first, to
delineate the properties of the dislocation core, which will require extensive
computation to unveil.  Dislocation cores play a significant role in finite
sized systems; in experiments to date, core radii may be only an order of
magnitude smaller than the system size.  Also, one needs to take into account
a possible transition to a hexatic phase, and then the transition of the
hexatic to a vortex liquid.  The latter process involves the unbinding of the
dislocations into disclination pairs, whose description requires the energy
and size of the dislocation core \cite{HalNel}.

    Dislocations in non-uniform condensates present further unsolved problems,
analogous to those for vortices in inhomogeneous systems, for example, the
dependence of the energy of a dislocation on position.  This dependence is
needed to determine whether it is energetically preferable for dislocations to
form at the edge, or by unbinding in the bulk of the condensate.

    We thank Markus Holzmann for helpful discussions and Joshua Rubin. This work 
was supported in part by NSF Grants PHY03-55014, PHY05-00914, and 
PHY07-01611. \\ \\

\section*{APPENDIX}

    In this Appendix we derive the velocity and displacement fields around an
elastic dislocation in a homogeneous two-dimensional vortex lattice, in
particular, for superfluid vortex lattices, a two-dimensional edge dislocation
\cite{Nab}.  The challenge is to find a solution of Eqs.~(\ref{veldisrel})
and ~(\ref{main1}) that satisfies Eq.\,(\ref{burgerscond}) for the Burger's
vector.  We first recast Eq.\,(\ref{burgerscond}) as
\begin{equation}
\label{burgerscond2}
\varepsilon_{ij}\frac{\partial w_{jk}}{\partial x_i} =
 b_k\delta(\ve{r}-\ve{r}'),
\end{equation}
where $\varepsilon_{ij}$ is the anti-symmetric tensor, and $w_{ij} =
\partial\epsilon_j/\partial x_i$.  Note that $\ve{\epsilon}$ is not continuous
within the entire plane (i.e., $\partial^2 \ve{\epsilon}/\partial x\partial y
\ne \partial^2 \ve{\epsilon}/\partial y\partial x$); otherwise,
Eq.\,(\ref{burgerscond2}) is not valid.

    To satisfy the conditions, $\ve{\nabla}\cdot\ve{v}=0$ and $\ve{\sigma} +
2mn\cprod{\ve{\Omega}}{\ve{v}}=0$, we define two dual (gauge) fields, the
stream function, $\psi$, and a modified Airy stress function, $\chi$.  Using
the elastic stress tensor, $\sigma_{ij}$, defined from Eq.\,(\ref{elforce})
such that $\sigma_j = -\partial \sigma_{ij}/\partial x_i$, we write,
\begin{eqnarray}
\label{streamf}
v_i &=& \varepsilon_{ij}\frac{\partial\psi}{\partial x_j}
\\
\label{airyf}
\sigma_{ij} &=&
 \varepsilon_{ik}\varepsilon_{jl}\frac{\partial^2\chi}{\partial x_k\partial
  x_l} -
2mn\Omega\delta_{ij}\psi.
\end{eqnarray}
Substituting these expressions into the curl of
Eq.\,(\ref{veldisrel}) and Eq.\,(\ref{burgerscond2}), we have,
\begin{eqnarray}
\label{dualdis}
 \frac{2C_1+C_2}{16C_1C_2}\nabla^4\chi - \frac{mn\Omega}{4C_1}\nabla^2\psi
  &=& (\cprod{\ve{\nabla}}{\ve{b}}\,\delta(\ve{r}-\ve{r}'))_z\hspace{0.2in}
  \\
\label{dualother}
\frac{\Omega}{4C_1}\nabla^2\chi + \nabla^2\psi -
\frac{mn\Omega^2}{C_1}\psi &=& 0,
\end{eqnarray}
where $\ve{r}'$ is the position of the dislocation.  Taking $\ve{r}'=0$,
we find
\begin{widetext}
\begin{eqnarray}
\label{chi} \chi(\ve{r}) &=&
\varepsilon_{ij}b_j\frac{\partial}{\partial
 x_i}
\left[\frac{C_2}{\pi}\modulus{\ve{r}}^2\left(\ln\left(
\frac{\modulus{\ve{r}}}{a}\right)-1\right)+
\frac{4C_2^2\delta^2}{(2C_1+C_2)\pi}\left(\ln\modulus{\ve{r}}
 +K_0\!\left(\frac{\modulus{\ve{r}}}{\delta}\right)\right)\right]\\
\label{psi}
\psi(\ve{r}) &=&
 \varepsilon_{ij}b_j\frac{\partial}{\partial
 x_i}\left[\frac{2C_2\Omega\delta^2}{(2C_1+C_2)\pi}\left(\ln\modulus{\ve{r}}
 +K_0\!\left(\frac{\modulus{\ve{r}}}{\delta}\right)\right)\right].
\end{eqnarray}
\end{widetext}

    The Airy stress function contains two basic terms.  For finite $\delta$,
the first term in Eq.~(\ref{chi}) is long ranged while the second term dies
off.  The stream function is proportional to this second term.
For imaginary $\delta$ we replace $K_0(x)$ by $-\frac{\pi}{2}Y_0(x)$ (where
$Y_0$ is a modified Bessel function of the first kind) since $\chi(\ve{r})$
and $\psi(\ve{r})$ are real.  [For $\delta\to\infty$, we find the typical
results for an edge dislocation in two dimensions (see Ref~\cite{Landau}),
however, this limit is not physically relevant here.]

\subsection{Energy of a Single Dislocation}

\begin{figure}[htbp]
\begin{center}
\vspace{0.25in}
{\includegraphics{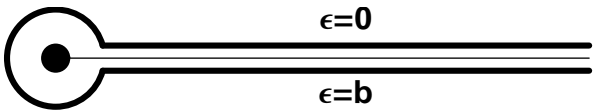}}
\vspace{0.25in}
\caption{Branch cut extending from the core of a
 dislocation.\label{fig:branchcut}}
\end{center}
\end{figure}

    According to Eq.~(\ref{burgerscond}), the displacement field,
$\ve{\epsilon}$, along any closed path around a dislocation does not return to
its initial value.  Instead, the displacement field at the close of the loop
is equal to its initial value plus the Burger's vector.  Thus, the system
includes a ``branch cut" from the dislocation core to the edge of the system,
as in Fig.~\ref{fig:branchcut}; the displacement field for a dislocation is
single valued.  (This structure is similar to that of the phase, $\phi$,
around a vortex, which changes from $0$ to $2\pi$; jumping by $2\pi$ across
the branch cut.)

    The energy of an elastic displacement in the superfluid vortex lattice is,
\begin{eqnarray}
\label{vlenergy}
E_{\rm el} &=&
  \int \left[2C_1({\ve \nabla}\cdot{\ve\epsilon})^2 +
  \frac12 nm\modulus{\ve{v}_{\rm T}}^2\right.
  \\
  &+& \left.C_2\left[\left(\frac{\partial \epsilon_x}{\partial x}
 -\frac{\partial\epsilon_y}{\partial y}\right)^2 +
 \left(\frac{\partial \epsilon_x}{\partial y} +
 \frac{\partial
  \epsilon_y}{\partial x}\right)^2\right]\right] d^2r,
  \nonumber
\end{eqnarray}
where $\ve{v}_{\rm T}$ is the transverse flow velocity.  Integrating by
parts and using $\modulus{\ve{v}_{\rm T}}^2 =
2\ve{\epsilon}\cdot\cprod{\ve{\Omega}}{\ve{v}}$, from Eq.~(\ref{veldisrel}),
we have
\begin{eqnarray}
\label{energystuff} E_{\rm el} &=&
\oint \left[2C_1\frac{\partial \epsilon_i}{\partial
 x_i}\epsilon_j\right.
\\ &&\,+ \left. C_2 \left(\left(\frac{\partial \epsilon_i}{\partial
 x_j} + \frac{\partial \epsilon_j}{\partial x_i}\right)\epsilon_i
- (\ve{\nabla}\cdot\ve{\epsilon})\epsilon_j \right)
 \right]d\Sigma_j \nonumber
\\
&+& \int \ve{\epsilon}\cdot\left[-2C_1\ve\nabla(\ve\nabla\cdot\ve\epsilon) -
C_2\nabla^2\ve\epsilon +
 nm\cprod{\ve{\Omega}}{\ve{v}}\right]d^2r\nonumber,
\end{eqnarray}
where $\Sigma_j$ is the normal component of the surface element of the
section of the boundary (including the branch cut).  In the absence of
external forces, Eq.\,(\ref{main1}) implies that the integrand in the second
integral in Eq.\,(\ref{energystuff}) is zero.  The integrals at the boundary
of the system and around the core are negligible; the branch cut contribution
is dominant.  Since $\ve{\epsilon}=\ve{b}$ on the lower part of the branch
cut, and $\ve{\epsilon}=0$ on the upper part (Fig.~\ref{fig:branchcut}),
we find finally
\begin{eqnarray}
E_{\rm el} &=&
\label{sdenergyb}
\int \left[2C_1b_j(\ve{\nabla}\cdot\ve{\epsilon})\right.
\\
\nonumber
&+& \left.C_2\left(b_i\frac{\partial \epsilon_i}{\partial x_j} +
b_i\frac{\partial \epsilon_j}{\partial x_i} -
b_j(\ve{\nabla}\cdot\ve{\epsilon}) \right)\right]d\Sigma_j^-,
\end{eqnarray}
where the integral is along the lower branch.  Computing this expression
using Eqs.~(\ref{streamf}) and (\ref{airyf}), relating the derivatives of $\ve{\epsilon}$ to $\sigma_{ij}$ \cite{stressf}, and (\ref{chi}) and (\ref{psi}), we
obtain Eq.~(\ref{sdenergy}).\\

\subsection{Energy of Many Interacting Dislocations}

    To obtain a more realistic description of the system, requires taking into
account the interactions of dislocations.  To compute the energy of a
two-dimensional gas of interacting dislocations we first substitute
Eq.\,(\ref{streamf}) and Eq.\,(\ref{airyf}) into Eq.\,(\ref{vlenergy}) by relating the derivatives of $\ve{\epsilon}$ to $\sigma_{ij}$ \cite{stressf}, and
find,
\begin{eqnarray}
E_{\rm el} &=&
 \int \Bigg[\frac{1}{8C_2}\left(\frac{\partial^2\chi}{\partial x_i\partial
 x_j}\right)^2 -
 \frac{2C_1-C_2}{32C_1C_2}(\nabla^2\chi)^2
\\
\nonumber
\label{elasticenergy2}
&-& \frac{mn\Omega}{4C_1}\psi\nabla^2\chi +
\frac{(mn\Omega)^2}{2C_1}\psi^2 +
\frac12mn\modulus{\ve{\nabla}\psi}^2\Bigg]  d^2r.
\end{eqnarray}

    We consider an aggregate of dislocations with Burger's vectors,
$\ve{b}_{\alpha}(\ve{r}_{\alpha})$, such that the sum over all Burger's
vectors is zero and $\chi(\ve{r})$ and its derivatives are zero at infinity --
unlike a single dislocation, a pair of dislocations with equal and opposite
Burger's vectors is not long ranged.  Integrating Eq.\,(\ref{elasticenergy2})
by parts we obtain,
\begin{equation}
 E_{\rm el} = \frac12\int\chi\left(\frac{2C_1+C_2}{16C_1C_2}\nabla^4\chi -
 \frac{mn\Omega}{4C_1}\nabla^2\psi\right) d^2r,
\end{equation}
and substituting in Eq.\,(\ref{dualdis}) and Eq.\,(\ref{chi}), and
integrating, we derive Eq.~(\ref{interdisE}).


\begin{thebibliography}{99}
\bibitem{lame} The Lam\'e coefficients, $\mu$ and $\lambda$, used in
\cite{HalNel}, are related to the moduli $C_1$ and $C_2$ in two dimensions by
$\mu = 2C_2$ and $\mu + \lambda = 4C_1$.
\bibitem{stressf} The elastic stress force, $\ve{\sigma}$, is related to
the usual elastic stress tensor, $\sigma_{ij} = 4C_2 u_{ij} +
(4C_1-2C_2)u_{kk}\delta_{ij}$, by $\sigma_i = \partial\sigma_{ij}/\partial
x_j$, where $u_{ij} = (\partial\epsilon_i/\partial x_j +
\partial\epsilon_j/\partial x_i)/2$ \cite{Landau}.
\bibitem{stable} Since $2C_1+C_2\le0$ in (and near) the incompressible
limit, $G_{\rm LL}$ changes sign for $k^2\ge-2\Omega^2nm/(2C_1+C_2)\simeq
16/\ell^2$, which according to Eq.~(\ref{gfe}) should be accompanied by a
resonant $\omega^2$ becoming negative.  This change of sign does not imply an
instability of the lattice, since it occurs for wavelengths $\lesssim$ the
vortex lattice spacing, a length scale at which the elastohydrodynamic
description is no longer valid.  In general, a conventional elastic system is
unstable if $C_1<0$ \cite{Landau}; however, in the vortex case the kinetic
energy term in Eq.~(\ref{eden}) guarantees the stability of the lattice.
\bibitem{delta} For $\delta^2>0$, the elastic Green's function of the
vortex lattice decays exponentially, while, if $\delta^2<0$, then the
penetration depth is imaginary, suggesting ripples in the displacement field.
In the regime where $\delta^2$ is negative, the quasi-wavelengths of any
undulations in the displacement are $\lesssim\ell/4$.
\bibitem{deltadep} The dependence of $T_{\rm m}$ on $\delta$ is negligible
in the region $\delta \lesssim a_{\rm c}$.  If one assumes incompressibility,
$C_{1,{\rm R}}\to\infty$, as in the derivation of $T_{\rm m}$ in
Ref.~\cite{HubDon}, for which $\delta$ becomes effectively infinite, the
melting temperature is given by the limit of $(a^2/\pi)C_{1,{\rm R}}C_{2,{\rm
R}}/(2C_{1,{\rm R}}+C_{2,{\rm R}})$, where the $C$'s are the renormalized
elastic moduli.  However, this limit is unphysical, since the Tkachenko mode
frequencies are finite only if $C_{1,{\rm R}}$ is finite.
\bibitem{extdis} In Sec.~\ref{sec:equ}, we separated out the elastic force
due to an inhomogeneous density, $\ve{\zeta}_n = \gamma n_v^0 \ve\nabla n$,
and treated it as an external stress.  This force only affects the
displacement field of the vortex lattice, according to Eq.~(\ref{rad-dis}),
and does not influence the displacement field of the dislocation.  However,
the coupling between the particle density and the vortex density  contributes
a term to the dislocation energy,
\begin{eqnarray}
\Delta E &=& \int \gamma n(\ve{r})\delta n_{\rm v} d^2 r = -\int \gamma
n(\ve{r})n_{\rm v}^0\ve{\nabla}\cdot\ve{\epsilon}\,d^2 r.  \nonumber
\end{eqnarray}
Using Eqs.~(\ref{veldisrel}), (\ref{streamf}), and (\ref{psi}), we find
for a vortex at $\ve{r}_0$,
\begin{equation}
\Delta E = \alpha\int
 n(\ve{r})\frac{\cprod{\ve{b}}
 {(\ve{r}-\ve{r}_0)}}{\modulus{\ve{r}-\ve{r}_0}^2}
 \,K_1\!\left(\frac{\modulus{\ve{r}-\ve{r}_0}}{\delta}\right)d^2 r,
\nonumber
\end{equation}
where $\alpha = \gamma C_2 n_{\rm v}^0/(2C_1+C_2)\pi$.  But because
$\delta<\ell$, this term is small and does not contribute significantly to the
overall energy.
\end{thebibliography}
\end{document}